\documentclass[twocolumn,showpacs,preprintnumbers,amsmath,amssymb,prb,aps]{revtex4-1}
\usepackage{epsfig}
\usepackage{epstopdf}
\usepackage{graphicx}
\usepackage{dcolumn}
\usepackage{bm}
\usepackage{textcomp}
\usepackage{array}
\usepackage{caption}
\usepackage{xfrac}
\usepackage{mathtools}
\usepackage{amsmath}
\usepackage{subfigure}

\begin{document}

\title{\textit{PY-Nodes}: An \textit{ab-initio} python code for searching nodes in a material using \textit{Nelder-Mead's} simplex approach.}
\author{Vivek Pandey$^{1}$ }
\altaffiliation{vivek6422763@gmail.com}
\author{Sudhir K. Pandey$^{2}$}
\altaffiliation{sudhir@iitmandi.ac.in}
\affiliation{$^{1}$School of Physical Sciences, Indian Institute of Technology Mandi, Kamand - 175075, India\\
$^{2}$School of Mechanical and Materials Engineering, Indian Institute of Technology Mandi, Kamand - 175075, India}
\date{\today}

\begin{abstract}
     
     With the discovery of topological semimetals, it has been found that the band touching points near the Fermi level are of great importance. They give rise to many exciting phenomena in these materials. Moreover, these points, commonly known as nodes, are related to several properties of these semimetals. Thus, the proper estimation of their coordinates is extremely needed for better understanding of the properties of these materials. We have designed a Python 3 based code named \textit{PY-Nodes} for efficiently finding the nodes present in a given material using \textit{first-principle} approach. The present version of the code is interfaced with the WIEN2k package. For benchmarking the code, it has been tested on some famous materials that possess characteristic nodes. These include - TaAs, a well-known Weyl semimetal, Na$_3$Bi, which is categorized as Dirac semimetal, CaAgAs, classified as a nodal-line semimetal and YAuPb, which is claimed to be non-trivial topological semimetal. In the case of TaAs, 24 nodes are obtained from our calculations. On computing their chiralities, it is found that 12 pairs of nodes having equal and opposite chirality are obtained. Furthermore, for Na$_3$Bi, a pair of nodes are obtained on either side of the $\Gamma$-point in the $\boldsymbol{k_3}$ direction. In the case of CaAgAs, several nodes are obtained in the $k_z$=0 plane. These nodes, when plotted in the $k_x$-$k_y$ plane, form a closed loop which is generally referred to as a nodal-line. Finally, in the case of YAuPb, large number of nodes are obtained in the vicinity of $\Gamma$-point. The results obtained for these materials are in good match with the previous works carried out by different research groups. This assures the reliability and efficiency of the \textit{PY-Nodes} code for estimating the nodes present in a given material.\\\\
\textbf{Program summary -}\\
\textit{Program Title}: \textit{PY-Nodes}\\
\textit{Program Files doi}:\\
\textit{Licensing provisions}: GNU General Public License 3.0\\
\textit{Programming language}: Python 3\\
\textit{External routines/libraries}: Math, Time\\
\textit{Nature of problem}: Searching for the node points corresponding to any given number of bands present in the first Brillouin zone of any material.\\
\textit{Solution method}: \textit{Nelder-Mead's} simplex approach is a well-known function-minimization method. This approach is used to minimize the function $f(\boldsymbol k)$, which is defined as sum of the absolute energy difference of the adjacent pairs of bands at a given $\boldsymbol k$-point. The local minima will correspond to the node points present in the material.

\end{abstract}

\maketitle

\section{Introduction} 
\setlength{\parindent}{3em}
  In present days, the band touching points near the Fermi level are paid much attention in the topological analysis of the materials\cite{1,2}. This is because these points give rise to many exciting phenomena in the topological materials such as topological semimetals. These band touching points are popularly known as node points or in some cases, as nodal-lines\cite{Ashwin}. The node points are formed when two or more bands touch each other at distinct points. On the other hand, the nodal-line is generated when two or more bands touch each other at a large number of points which collectively form a closed loop\cite{Line-node1,Line-node2}. In the topological semimetals, the nodes with the non-trivial properties are formed between the bands coming from both the valence and conduction bands. These nodes are generally associated with many topological quantities such as chirality of nodes\cite{Dirac}, Berry curvature\cite{BerryCurvature}, surface states\cite{SurfaceState} etc. These properties greatly determine the behaviours of the semimetals.  
\par One of the famous classes of semimetals is Weyl semimetal\cite{WS1,WS2,WS3,WS4}. This class of materials possess either inversion symmetry or the time-reversal symmetry, but not both at a time\cite{Weyl-sym}. Furthermore, they are characterized by the presence of several nodes near the Fermi level. These are generally known as Weyl nodes. They play an important role in deciding the properties of the semimetal. In general, these nodes are associated with specific non-zero chiralities ($\textit{C}$)\cite{Weyl-chirality}. Based on the sign of the chirality of a given Weyl node, it acts as the source (+ve chiral nodes) or sink (-ve chiral nodes) of the Berry flux\cite{Berry-flux}. Thus, the chirality of a Weyl node is directly related to the Berry flux. Quantitatively, $\textit{C}$ of a Weyl node situated at point $\textbf{\textit{k}}_\textbf{0}$ is defined as $\frac{1}{2\pi}$ times the net Berry-flux penetrating through any surface enclosing $\textbf{\textit{k}}_\textbf{0}$. Also, in the case of Weyl semimetals, there is no net Berry-flux penetrating in or coming out of the first Brillouin zone (BZ). This generally suggests that Weyl nodes always occur in pairs of opposite chirality in the first BZ\cite{Nielsen}. In addition to this, it is commonly seen in the case of Weyl semimetals that the surface states enter into the bulk only through these node points. Furthermore, the number of nodes in the BZ is closely related to the symmetry present in the material. If there exists an inversion symmetry in a Weyl semimetal, then for every node of chirality $\textit{C}$ and situated at crystal momenta $\textbf{\textit{k}}$, there exists another node point of opposite chirality situated at -$\textbf{\textit{k}}$. Thus, for such Weyl semimetals, the total number of nodes in the BZ must be in multiples of two. However, if the Weyl semimetal possesses time-reversal symmetry then the Weyl nodes situated at the points $\textbf{\textit{k}}$ \& -$\textbf{\textit{k}}$ are of the same chirality. And therefore, to nullify these chiralities, there exists two other Weyl nodes at some other crystal momenta and having opposite chirality. Thus, the total number of Weyl nodes in this type of Weyl semimetal is in multiple of four\cite{Ashwin}.

\par There is another class of semimetals in which both the symmetries are present simultaneously. In these semimetals, there exists two degenerate Weyl nodes of opposite chirality at the same crystal momenta. This results in the four-fold degeneracy at these points. This class of semimetals is popularly known as Dirac semimetals. The degenerate points are usually called Dirac nodes and the cone-like structure formed near these points are commonly known as Dirac cones. These nodes are also found close to Fermi level. Unlike Weyl nodes, the Dirac nodes are neither symmetry protected nor do they possess a non-zero chirality. Apart from the Dirac and Weyl semimetals, there is yet another class of semimetals which are popularly known as nodal-line semimetals. In this class of semimetals, the topmost valence band (VB) and the bottommost conduction band (CB), instead of touching at a single point, touches in a closed region\cite{NS1,NS2,NS3,NS4}.

\par Most of the topological properties of these semimetals are associated with the band touching points (\textit{i.e.}, node points or the nodal-lines) near the Fermi level, in one or other manner. Thus, to have a greater insight of these materials, proper estimation of the coordinates of these band touching points becomes utmost necessary. In the present scenario, tight-binding (TB) based models are generally used to study the different topological properties of materials. These studies also include the estimation of the topological node points. At present, there are various well-known ways to construct the TB models. These include - (I) Slater-Koster method\cite{SK-M}, (II) Maximally localized Wannier function (MLWF)\cite{MLWF}, (III) discretization of the $\boldsymbol{k}.\boldsymbol{p}$ model onto a lattice\cite{kp-M}, etc. Among these, the MLWFs are most widely used to construct the TB model for the real material simulation. The process is implemented in Wannier90\cite{Wannier90} code which is interfaced with the WIEN2k\cite{WIEN2k} package. However, the wannierization procedure involved in generating the TB model is sensitive to a large number of parameters. Firstly, one must decide the energy window corresponding to which the TB model is required. A good wannier fitting requires that the bands in this energy window must be disentangled from the bands in the other energy regions. Also, major weight of the projectors contributing to these bands must fall in the energy window corresponding to which wannierization is required. Furthermore, this energy window must be as small as possible to reduce the computational cost. Thus, it becomes a challenging task to produce a good TB model corresponding to a given real system. But once the TB model is achieved, it proves to be very useful in calculating different topological properties, including the coordinates of the node points with an acceptable accuracy. However, in some complex systems, such as YAuPb, where the nodes are highly concentrated in a very small region\cite{Vivek}, the TB model may fail to estimate their accurate coordinates. Also, in the model-based analysis, there are possibilities of missing some of the node points present in the material. This is because the model only approximates the real system and is not the exact one. This generally demands that these band touching points must be estimated from the \textit{first-principle} approach instead of using the model-based calculations.

\par Nodes between a pair of bands can be mathematically visualized as the local minima of the energy gap function associated with the bands. Thus, in the computational codes such as WannierTools\cite{WannierTools}, nodes are usually searched by minimizing the energy gap function. In general, this task is achieved by employing various standard function-minimization algorithms. Some of these algorithms are \textit{Nelder and Mead’s Downhill Simplex Method}\cite{Nelder}, \textit{Conjugate Gradient Methods}\cite{Conjugate_Grad}, \textit{Quasi-Newton Method}\cite{Qu-Ne}, etc. These algorithms are generally based on simplex approaches. The term \textquotedblleft simplex" refers to the set of data points. Using these simplexes, the algorithms search the local minima in an iterative way. Each initial simplex provides one local minima. Finally, the algorithms search for nodes from these local minima. The \textit{Conjugate Gradient Method} uses the vector approach for the function-minimization. The method requires the storage of all previous searching directions and residue vectors. In addition to this, it also requires keeping track of a series of many matrix-vector multiplications. This makes the method computationally expensive. Moving further, the \textit{Quasi-Newton Method} is based on the matrix approach. It includes the calculations of Jacobian which is generally very costly in terms of computational time. It is also important to mention here that the \textit{Conjugate Gradient Method} and \textit{Quasi-Newton Method} involve the derivative of the function for which the minima is required. For the present case, this function will be the energy gap function. In order to find the derivative of a function, its functional form must be known. In the present case, the functional form of the energy gap function can be obtained using well-known interpolation methods such as Newton’s interpolation\cite{Sastry} or Lagrange interpolation\cite{Sastry}. However, there may be the cases where the bands, corresponding to which the energy gap function is required, are highly dispersive. That is, the bands are not smooth and change rapidly with the change in \textbf{\textit{k}}-points. For such bands, even after taking a large number of \textbf{\textit{k}}-points, one may not get the proper interpolated energy gap function. The reliability of the derivative function obtained using such an interpolated function will also become questionable. Furthermore, the interpolated function will be associated with a large number of coefficients which need to be stored. This will require higher storage capacity. Such issues with these methods make them inefficient for function-minimization process as per required for the present physical problem. Unlike these methods, the \textit{Nelder and Mead’s Downhill Simplex Method} involves the simple arithmetic operations. In each cycle, it carries out the reflection, contraction and expansion of one data point through the centroid of all other data points of the simplex. Also, to an extra advantage, it does not require to keep the records of the previous iterations. Moreover, it calculates the energy of bands and the energy gap value using the \textit{first-principle} method. Thus, the calculated energy gap is accurate and not the approximated value based on some interpolated function. So, it is the better choice for function-minimization as compared to other methods discussed above.

\par In the present work, a Python 3 based code named \textit{PY-Nodes} is designed for searching nodes associated with two or more bands in a given material using the \textit{first-principle} approach. The code is presently interfaced with the WIEN2k package\cite{WIEN2k}. The algorithm of the code is based on the \textit{Nelder-Mead's} function-minimization approach. For checking the reliability of the code, it is used to find out the nodes present in some well-known materials. These include - TaAs\cite{TaAs}, Na$_3$Bi\cite{Na3Bi}, CaAgAs\cite{CaAgAs} and YAuPb\cite{YAuPb_claim}. These materials belong to the class of Weyl semimetals, Dirac semimetals, nodal-line semimetals and topological semimetals, respectively. The coordinates of the nodes obtained from the \textit{PY-Nodes} code are also compared with the values reported in the literatures corresponding to each material.

\section{Theoretical Background}

\subsection{Nelder and Mead’s Downhill Simplex Method}
  The problem of searching the node points between \textit{n} bands can be visualized as finding the local minima of the function $f(\boldsymbol k)$ which is defined as the sum of the absolute energy difference of the adjacent pairs of bands at a given $\boldsymbol k$-point. As previously discussed, this can be effectively carried out using the \textit{Nelder and Mead’s Downhill Simplex Method}\cite{Nelder} which is implemented in the present code. In this method, to minimize a function of \textit{n} independent variables, the initial simplexes are required to be formed of (\textit{n}+2) data points. Also, the necessary condition to start the function-minimization process is that the values of the function at all the points of the initial simplex must not be simultaneously equal. During the minimization process, points in the initial simplex will evolve with the repeated application of the three processes, namely - reflection, contraction and expansion. These processes are carried out using the reflection coefficient ($\alpha$), contraction coefficient ($\beta$) and expansion coefficient ($\gamma$), respectively.
\par Let us discuss the method in further detail. Suppose we need to find the local minima of the function $f(\boldsymbol k)$. Consider the initial simplex of \textit{n} number of data points which are denoted as $\boldsymbol{k_1}$, $\boldsymbol{k_2}$, $\boldsymbol{k_3}$, ...., $\boldsymbol{k_n}$. Also, let the values of the function at these points are $y_1$, $y_2$, $y_3$, ...., $y_n$, respectively. It must be noted here that the \textit{n} number of independent variables in $f(\boldsymbol k)$ suggest that the initial simplex must be formed of (\textit{n}+2) data points. Now, at first the algorithm finds out the data points $\boldsymbol{k_l}$ \& $\boldsymbol{k_h}$ from the initial simplex corresponding to which the function $f(\boldsymbol k)$ has the lowest and the highest value, denoted as $y_l$ \& $y_h$, respectively. Nextly, the algorithm will find out the centroid of all the data points in the simplex excluding the $\boldsymbol{k_h}$. Let the centroid be denoted as $\boldsymbol{k_C}$ and the corresponding value of function be denoted as $y_C$. The algorithm proceeds with the reflection of the point $\boldsymbol{k_h}$ through the centroid \textit{i.e.}, $\boldsymbol{k_C}$ to find out a new data point, say $\boldsymbol{k^*}$. The value of $\boldsymbol{k^*}$ depends on the magnitude of reflection coefficient ($\alpha$) through the following relation\cite{Nelder},

\begin{equation}
  \boldsymbol{k^*}=(1+\alpha)\boldsymbol{k_C} - \alpha \boldsymbol{k_h}
\end{equation}
Here, $\alpha$ can take any positive value. If $y^*$=$f(\boldsymbol{k^*})$ is less than $y_l$, then it generally suggests that the value of the function is decreasing on moving in the direction of reflection. Thus, the algorithm further searches for another new point in the same direction by using the expansion coefficient ($\gamma$) through the following relation\cite{Nelder},
\begin{equation}
  \boldsymbol{k^{**}}=(1+\gamma)\boldsymbol{k^*} - \gamma \boldsymbol{k_C}
\end{equation}
The value of $\gamma$ must be greater than or equal to 1. Now, if $y^{**}$=$f(\boldsymbol{k^{**}})$ is less than $y_l$, the data point $\boldsymbol{k_h}$ is replaced by $\boldsymbol{k^{**}}$. It must be noted that the function has the lowest value at $\boldsymbol{k^{**}}$ at this stage. However, if $y^{**}$$>$$y_l$, then $\boldsymbol{k_h}$ is replaced by $\boldsymbol{k^{*}}$. Thus, a new simplex is obtained. There may be the case when the value of $y^*$ might not be the lowest value. Then the algorithm checks for the following condition,
\begin{equation}
  \boldsymbol{k^{*}}>\boldsymbol{k_i} , \hspace*{0.2in}\forall  i\ne h
\end{equation}
If equation 3 is not satisfied, then also the value of $\boldsymbol{k_h}$ is replaced by $\boldsymbol{k^{*}}$ to obtain a new simplex. However, if the above equation is satisfied, it generally suggests that the algorithm is searching the new points in the region away from the local minima. Thus, to move towards the local minima, the further mentioned steps are carried out. Firstly, the algorithm checks for the condition given below,
\begin{equation}
  \boldsymbol{k^{*}}>\boldsymbol{k_h}
\end{equation}
If equation 4 is not satisfied, the value of $\boldsymbol{k_h}$ is replaced by $\boldsymbol{k^{*}}$, else it remains unchanged. Having done this, a new data point is obtained by using the contraction coefficient ($\beta$) through the following relation\cite{Nelder},
\begin{equation}
  \boldsymbol{k^{**}}=\beta \boldsymbol{k_h} + (1-\beta) \boldsymbol{k_C}
\end{equation}
In the above equation, the coefficient $\beta$ must be assigned any value between 0 and 1. Now, if $y^{**}$=$f(\boldsymbol{k^{**}})$ is greater than $y_h$, then all the data points $\boldsymbol{k_i}$ are replaced by ($\boldsymbol{k_i}$+$\boldsymbol{k_l}$)/2. However, if $y^{**}$ is less than or equal to $y_h$, then $\boldsymbol{k_h}$ is replaced by $\boldsymbol{k^{**}}$ to get a new simplex. It is to be noted here that one time application of the above mentioned steps on the initial simplex will provide a new simplex. Furthermore, in comparison to the initial simplex, the data points in the new simplex will be closer to the local minima. This generally suggests that the iterative application of the above mentioned steps on the initial and the subsequent simplexes will be effective in searching the local minima. As the new data points are obtained from the old data sets, it seems obvious that for a given value of $\alpha$, $\beta$ \& $\gamma$, the obtained local minima will depend on the initial simplex. Thus, each initial simplex will correspond to one local minima.\\\\

\textit{Halting criteria:} It is very necessary to have an effective halting criteria for the above mentioned iterative method. This will be helpful in maintaining a good accuracy in the final coordinates of the local minima. In the above methodology, after the enough number of iterations, the values of all the $y_i$ corresponding to the data points of the simplex are expected to be very close to the real minima and also to one-another. This generally suggests that after a good number of iterations, the values of $y_i$ will be highly precise. Thus, the standard deviation of the $y_i$s will be extremely small. Hence, after every iteration, the standard deviation of $y_i$s is compared with a predefined limit. If the standard deviation becomes less than the preset limit, the iterative process stops. In this way a predefined limit for standard deviation of $y_i$s serves as the halting parameter for the process.

\subsection{Simplex Formation}
The code \textit{PY-Nodes} employs the \textit{Nelder and Mead’s} method to minimize the function $f(\textbf{k})$ which gives the value of sum of the absolute energy gap between the adjacent pairs of bands at a given $\textbf{\textit{k}}$-point. Through this approach, it successfully finds out the nodes associated with the given number of bands in the Brillouin zone (BZ) of a given crystal system. The function-minimization is carried out using a separate module named \textit{Nelder-Mead} provided with the \textit{PY-Nodes} code. The efficiency of the code mainly depends on two factors- (I) effective formations of initial simplexes and (II) an optimized values of $\alpha$, $\beta$ \& $\gamma$. The effective formations of simplexes mean that a number of small simplexes must be formed homogeneously in all the regions of the BZ. Furthermore, the values of $\alpha$, $\beta$ \& $\gamma$ should be such that the algorithm should efficiently search the new points only in the region close to the initial simplex. As already seen, the reflection, contraction and expansion operations always involve $\boldsymbol{k_h}$ and/or $\boldsymbol{k_C}$. For the efficient searching of nodes, the new data point generated upon any of these operations must not be too close to $\boldsymbol{k_h}$ or $\boldsymbol{k_C}$. Also, it must not be too far from $\boldsymbol{k_h}$ and $\boldsymbol{k_C}$. This is because, if the new points are generated too close to $\boldsymbol{k_h}$ or $\boldsymbol{k_C}$, it will take a very large number of iterations to converge the simplex to a local minima. Furthermore, if the new points are situated too far from $\boldsymbol{k_h}$ and $\boldsymbol{k_C}$, there is a high chance of missing the local minima. This may also result in the oscillations of the new generated points around the local minima. Thus, it will make the nodes-searching algorithm less efficient. Also, it can be seen from equation 1 that for very small values of $\alpha$, the new data point will be generated close to $\boldsymbol{k_C}$. For this case, as can be seen from equation 2, the expansion operation will become less effective. In addition to this, very high values of $\gamma$ will result in getting new data points which are situated too far from initial simplex. This may result in missing some regions of the BZ from being analysed for the presence of nodes. Apart from this, as can be seen from equation 5, for high values of $\beta$, the contraction process will result in getting new data points close to $\boldsymbol{k_h}$. Also, very small values of $\beta$ will result in getting the data point close to $\boldsymbol{k_C}$. From the above discussion, it can be concluded that these cases will make the nodes-searching process less efficient. Thus, optimized values of $\alpha$, $\beta$ \& $\gamma$ are needed for better performance of the code. In order to obtain the optimized values of these coefficients, calculations have been performed using different sets of $\alpha$, $\beta$ \& $\gamma$. Based on the time taken and the results obtained from these calculations, the values $\alpha$=0.6, $\beta$=0.5 \& $\gamma$=1 are suggested to be the optimized values and are expected to serve the purpose in most of the cases. However, one may also obtain another efficient set of values for $\alpha$, $\beta$ \& $\gamma$.

\par It is important to note that $f$ is the function of three independent variables namely, $\boldsymbol{k_x}$, $\boldsymbol{k_y}$ \& $\boldsymbol{k_z}$. As discussed before, for the function of \textit{n} independent variables, the simplex must be formed of (\textit{n}+2) data points. So, in the present case simplex must be formed from 5 data points. It is well-known that the \textit{\textbf{k}}-points sampling done in most of the computational codes are homogeneous in the BZ. Thus, if small simplexes are formed associated with each \textit{\textbf{k}}-point then they can be considered as effective. In the \textit{PY-Nodes} code, for every point ($k_x,k_y,k_z$) denoted as $P$, four other points are generated based on the \textit{shift} parameter. Out of these four points, three points are ($k_x$+\textit{shift}$,k_y,k_z$), ($k_x,k_y$+\textit{shift}$,k_z$) and ($k_x,k_y,k_z$+\textit{shift}), which are denoted as $M_1$, $M_2$ and $M_3$, respectively. The fourth point, denoted by $R$, is obtained by reflecting the original point through the centroid of the points $M_1$, $M_2$ and $M_3$. Let the centroid of $M_1$, $M_2$ and $M_3$ be denoted by $C$. The coordinates of $C$ will be given by ($k_x$+(\textit{shift}/3)$,k_y$+(\textit{shift}/3)$,k_z$+(\textit{shift}/3)). Thus, the coordinates of the point $R$ are given by ($k_x$+(2\textit{shift}/3)$,k_y$+(2\textit{shift}/3)$,k_z$+(2\textit{shift}/3)). Thus, for every \textit{\textbf{k}}-point $P$, the simplex will be formed of points $P$, $M_1$, $M_2$, $M_3$ and $R$. Also, the size of all the simplexes formed across the BZ is maintained to be uniform and can be changed by changing the value of \textquoteleft \textit{shift}' parameter in the input file (\textit{PY-Nodes.input}). Furthermore, to find all the nodes efficiently, the analysis of each and every region of the full BZ is required. The uniform size of simplexes and the optimized values of $\alpha$, $\beta$ \& $\gamma$ are expected to serve this purpose.

\subsection{Nodes in first Brillouin Zone}
Solids are highly symmetric\cite{Cracknel}. Thus, in order to study their electronic structures and other properties, only the first BZ of the systems are generally considered. The results corresponding to the first BZ are enough to comment on the properties of the solid as a whole. Therefore, to study different properties related to the nodes, it becomes necessary to find out the total number of nodes in the first BZ. However, the \textit{Nelder and Mead's} methodology does not guarantee that the nodes obtained will be confined in the first BZ. Thus, the final treatment of the nodes searched by the methodology becomes very necessary. This is to make sure that all the points are situated in the first BZ. To serve this purpose, the \textit{PY-Nodes} code is provided with a separate module named \textit{NodesFirstBZ}. The code \textit{NodesFirstBZ} will find out the lattice structure of the material from the \textit{struct\_num} parameter in the input file. It must be mentioned here that before the role of \textit{NodesFirstBZ} code, the coordinates of nodes are obtained in terms of conventional lattice vectors, \textit{i.e.} \boldsymbol{$k_x$}, \boldsymbol{$k_y$} \& \boldsymbol{$k_z$}. Having known the crystal structure of the material, the code converts the conventional coordinates of all the nodes in terms of primitive lattice vectors, \textit{i.e.} \boldsymbol{$k_1$}, \boldsymbol{$k_2$} \& \boldsymbol{$k_3$}. After this step, the code will check if a given node is in the first BZ or not. Furthermore, if any of the nodes obtained are outside the boundaries of  first BZ, the code will find out the coordinates of corresponding node in the first BZ. Also, at the time of writing the final result, the code makes sure that no node points are repeated in the output file \textit{i.e.}, \textit{nodes\_final.dat}. The method used to map all the nodes in the first BZ is discussed next.

\begin{table*}
\caption{\label{tab:table1}%
\normalsize{The details of various input parameters for \textit{PY-Nodes} code.
}}
\begin{ruledtabular}
\begin{tabular}{ccc}
\textrm{\textbf{Name}}&
\textrm{\textbf{Default value}}&
\textrm{\textbf{Meaning}}\\
\colrule
    \textit{case} & - &  WIEN2k self-consistently energy converged file name.\\
    \textit{struct\_num} & - &  Crystal structure number. (Refer Table II)\\
    \textit{alpha} & 0.6 &  Reflection coefficient.\\ 
    \textit{beta} & 0.5 &  Contraction coefficient.\\ 
    \textit{gamma} & 1.0 &  Expansion coefficient.\\ 
    \textit{SOC }(y/n) & - &  Spin-orbit coupling is included (y) or not (n).\\ 
    \textit{shift} & 0.2 &  Shifting parameter.\\ 
    \textit{coordinates\_prec} & 4 &  Precision limit in the coordinates of the nodes.\\ 
    \textit{halt\_conv\_limit} & $10^{-11}$ &  The convergence limit for the \textit{Nelder-Mead's} iterative method.\\ 
    \textit{band\_indices} & - &  Band indices corresponding to which nodes are required.\\ 
   \textit{ write\_lim} & $10^{-4}$ &  For a given initial set of points, the details of all the further generated points will be  \\
     &   & written in \textit{kpoints\_evolv.dat} file, if the function value of any data point becomes less than this limit.\\ 
    
\end{tabular}
\end{ruledtabular}
\end{table*}

\begin{table*}
\caption{\label{tab:table1}%
\normalsize{The structure number assigned to different crystal structures.
}}
\begin{ruledtabular}
\begin{tabular}{cc}
\textrm{\textbf{Crystal Structure}}&
\textrm{\textbf{Structure Number}}\\
\colrule
          Cubic Primitive           & 1\\
          Cubic face-centred        & 2\\
          Cubic body-centred        & 3\\
          Tetragonal Primitive      & 4\\
          Tetragonal body-centred   & 5\\
          Hexagonal Primitive       & 6\\
          Orthorhombic Primitive    & 7\\
          Orthorhombic base-centred & 8\\
          Orthorhombic body-centred & 9\\
          Orthorhombic face-centred & 10\\
          Anything else             & 0\\
    
\end{tabular}
\end{ruledtabular}
\end{table*}

Consider a lattice point (say $\Gamma$-point). The coordinates of the neighbouring lattice points corresponding to $\Gamma$-point will be given by $P_i$ ($p_1$,$p_2$,$p_3$), where $p_1$, $p_2$ \& $p_3$ are either $\pm$1 or 0. The coordinates are in the \boldsymbol{$k_1$}, \boldsymbol{$k_2$} \& \boldsymbol{$k_3$} basis. Note that all $p_j$ for any $P_i$ must not be simultaneously 0, as it will correspond to $\Gamma$-point. Thus, there will be 26 $P_i$ points. In the WIEN2k package, the \textbf{\textit{k}}-points are homogeneously sampled in the first BZ. Thus, all the simplexes formed will also be either in the first BZ or very close to its boundaries. Now, let the parameters \textit{alpha}, \textit{beta} and \textit{gamma} are set to their default values as mentioned in Table I. Then, it is highly unlikely that the nodes obtained by \textit{Nelder and Mead’s} method will be outside the volume enclosed by the $P_i$ points. The nodes which fall within the volume enclosed by the $P_i$ points can easily be verified if they are situated in the first BZ or not. Also, if any node is outside the first BZ but within the volume enclosed by the $P_i$ points, then its corresponding point in the first BZ can also be easily obtained. If a random point $A$ ($a_1$,$a_2$,$a_3$) is within the first BZ, then it will be closest to the $\Gamma$-point in comparison to its distance from all other $P_i$ points. By this criterion, if point $A$ is outside the first BZ, then the corresponding point in the first BZ will be $A'$ ($a_1-p_l$,$a_2-p_m$,$a_3-p_n$). Here, $P$($p_l$,$p_m$,$p_n$) is the point which is closest to $A$. With this mechanism, the code \textit{NodesFirstBZ} will find out all the nodes corresponding to the first BZ. At last, the code will also convert the coordinates of these nodes in the \boldsymbol{$k_x$}, \boldsymbol{$k_y$} \& \boldsymbol{$k_z$} basis, for non-primitive structures. It must be noted that there is less possibility of obtaining a node point outside the volume enclosed by the $P_i$ points. However, it may happen in some cases. For such node points, the above mentioned approach will not serve its purpose. Thus, extra care must be taken for finding all the nodes in the first BZ.

\par In the present version of the \textit{PY-Nodes} code, the module \textit{NodesFirstBZ} is capable of performing its task only for the structures explicitly mentioned in Table II (Structure Number: 1-10). For any other structure, the coordinates of nodes will only be mentioned in terms of \boldsymbol{$k_x$}, \boldsymbol{$k_y$} \& \boldsymbol{$k_z$} basis. Also, for any other structure, we do not claim that the nodes obtained will be inside the first BZ.

\begin{figure}[tbh]
  \begin{center}
    \includegraphics[width=0.95\linewidth, height=8.4cm]{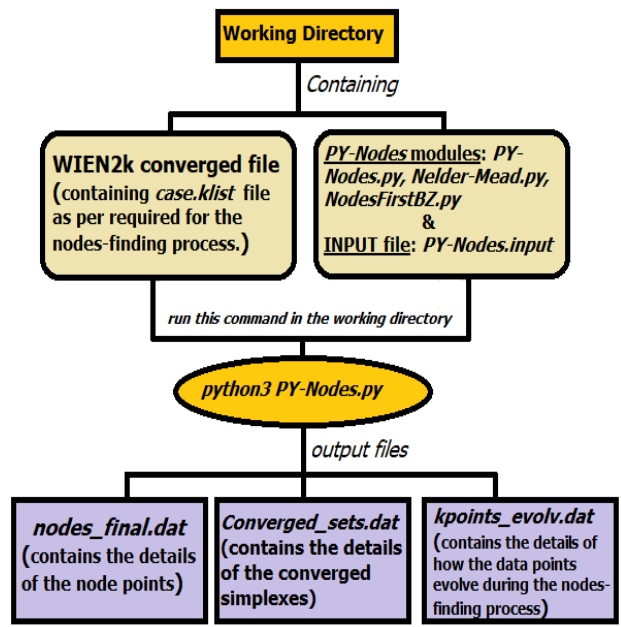} 
    \caption{\small Workflow of the \textit{PY-Nodes} code.}
    \label{fig:}
  \end{center}
\end{figure}

\section{Workflow and technical details}

\subsection{Workflow}
\par The workflow of the \textit{PY-Nodes} code is diagrammatically represented in Fig. 1. For using the code, user needs to keep the \textit{case} directory along with all the modules of the code and the input file (\textit{PY-Nodes.input}) in a separate directory. The details of different input parameters along with their suggested default values are mentioned in Table I. The \textit{case} directory stands for the directory containing self-consistently converged calculation of ground state energy of the given material, carried out using WIEN2k package. The user needs to explicitly mention in the input file if the spin-orbit coupling (SOC) is included in self-consistent calculation or not. For this purpose, the value of \textquotedblleft \textit{SOC}\textquotedblright \hspace*{0.05in}parameter in the input file can be set as \textquoteleft y', if SOC is included (otherwise \textquotedblleft \textit{SOC}\textquotedblright =\textquoteleft n'). In addition to this, the information about the structure of the material under study must be mentioned. This can be done by assigning the structure number from Table II to the \textit{struct\_num} parameter in the input file. Furthermore, for deciding the size of simplex, an appropriate value to the \textit{shift} parameter must be assigned. The role of \textit{shift} parameter has been already discussed in section II B. Nextly, the values of reflection, contraction and expansion coefficients must be provided by assigning respective values to the \textit{alpha}, \textit{beta} \& \textit{gamma} parameters in the input file. Furthermore, precision in the coordinates of the nodes can also be controlled by the user through the parameter named \textit{coordinates\_prec} in the input file. It is advised to keep the value of \textit{coordinates\_prec} less than 11. For higher values, there may be loss in precision in the final output and also, it may cause the improper format of the output. Moving further, the halting convergence limit for the \textit{Nelder-Mead's} method must be assigned to the parameter named \textit{halt\_conv\_limit}. The role of halting criteria has been also already discussed in section II A. Besides, the choice of the number of bands corresponding to which node points are required is left over to the users. For this purpose, band indices, separated by semicolon, must be provided, to the parameter named \textit{band\_indices} in the \textit{PY-Nodes.input} file. Here, one must note that suppose it is required to find out the \textbf{\textit{k}}-point where \textit{m} number of bands touches. Then, band indices of all these \textit{m} bands, separated by semicolon, must be provided in the \textit{band\_indices} parameter of the input file. Furthermore, after enough number of iterations, the value of $f(\boldsymbol k)$ at the data points in a simplex will become very small. This generally indicates that the simplex has reached very close to the local minima. At this stage it will be interesting to keep track of the information of how the data points further evolve during the process till the local minima is obtained. These data points may be used for plotting the band structure near the nodes. For this purpose, the value of $f(\boldsymbol k)$ at the data points is checked with the preset limit assigned to the parameter named \textit{write\_lim} in the input file. Once the value of $f(\boldsymbol k)$ at any data point becomes less than the value of \textit{write\_lim}, the information of all the further generated data points will be written in the \textit{kpoints\_evolv.dat} file. It must be noted here that only those simplexes are considered to contain the node points in which the value of $f(\boldsymbol k)$ at any data point becomes less than the value of \textit{write\_lim}. For such simplexes, after the iterative process stops, the data points with lowest value of $f(\boldsymbol k)$ are considered as node points. The details of all the node points will be written in a separate file named \textit{nodes\_final.dat}.

\begin{table*}
\caption{\label{tab:table1}%
\normalsize{The various input details and the \textit{k}-mesh size used to calculate the ground state energy of the materials using WIEN2k package. The \textit{k}-meshes are taken in the irreducible part of the Brillouin zone (IBZ).
}}
\begin{ruledtabular}
\begin{tabular}{ccccc}
\textrm{$\textbf{case}$}&
\textrm{$\textbf{space-group}$}&
\textrm{$\textbf{lattice parameters}$}&
\textrm{$\textbf{Wyckoff Positions}$}&
\textrm{\textbf{\textit{k}-mesh (in IBZ)}}\\
\colrule\\
    TaAs\cite{TaAs-lp} &$I4_1md$   & a=b=3.4824 \AA, c=11.8038 \AA      & Ta = (0.00, 0.00, 0.00)            & 10$\times$10$\times$10 \\
         &           & $\alpha$=$\beta$=$\gamma$=90       & As = (0.00, 0.00, 0.4176)          &                       \\\\
Na$_3$Bi\cite{Na3Bi-lp} &$P6_3/mmc$ & a=b=5.448 \AA, c=9.655 \AA       & Na (I) = (0.00, 0.00, 0.25)        & 12$\times$12$\times$6  \\
         &           & $\alpha$=$\beta$= 90, $\gamma$=120 & Na (II) = (1/3, 2/3, 0.583)        &                         \\
         &           &                                    & Bi = (1/3, 2/3, 0.25)              &                         \\\\
   YAuPb\cite{YAuPb-lp} & $F-43m$   & a=b=c=6.729 \AA                    & Y  = (0.25, 0.25, 0.25)            & 10$\times$10$\times$10 \\
         &           & $\alpha$=$\beta$=$\gamma$=90       & Au = (0.00, 0.00, 0.00)            &                       \\
         &           &                                    & Pb = (0.75, 0.75, 0.75)            &                       \\\\
  CaAgAs\cite{CaAgAs-lp} & $P-62m$   & a=b=7.204 \AA, c=4.27 \AA       & Ca = (0.5854, 0.00, 0.50)          & 8$\times$8$\times$12  \\
         &           & $\alpha$=$\beta$= 90, $\gamma$=120 & Ag = (0.2492, 0.00, 0.00)          &                         \\
         &           &                                    & As (I) = (1/3. 2/3, 0.00)          &                         \\
         &           &                                    & As (II) = (0.00, 0.00, 0.50)       &                       \\\\
\end{tabular}
\end{ruledtabular}
\end{table*}

\par To run the code, firstly one needs to move inside the \textit{case} directory. Then the \textit{case.klist} file must be generated using \textit{x kgen} (irreducible BZ) or \textit{x kgen -fbz} (full BZ) command. The \textit{case.klist} file defines the number of \textbf{\textit{k}}-points corresponding to which simplex must be formed and the nodes must be searched. Having done this, one needs to come out of the \textit{case} directory to the main directory and run the following command,\\\\
\hspace{0.8 cm} \hspace{0.8 cm} \textit{python3 PY-Nodes.py}\\

The final results will be mentioned in the file named \textit{nodes\_final.dat}. Furthermore, the details about the converged simplexes will be mentioned in the file named \textit{Converged\_sets.dat}. Apart from this, the details about how the \textbf{\textit{k}}-points evolved corresponding to each node point will be mentioned in the \textit{kpoints\_evolv.dat} file. 

The number of nodes obtained depends on the number of $\textbf{\textit{k}}$-points corresponding to which the calculations have been carried out. If a sufficiently large number of $\textbf{\textit{k}}$-points are not taken then one may not obtain all the nodes present in a given material. To make sure that all the nodes present in a given material have been obtained, one can check the convergence of the number of nodes obtained with the increase in number of $\textbf{\textit{k}}$-points. Furthermore, to search each and every region of the first BZ for the presence of node points, a very large size of \textit{k}-mesh is needed. This may be very costly in terms of computational time. Instead of following this approach, we suggest an efficient method to find out all the nodes present in a given material. One must keep running the code on different sizes of small \textit{k}-meshes. Also, keep storing the information of all the nodes obtained from these calculations in a single file. Care must be taken while storing the details of the nodes so that no nodes are repeated. After running the code on a certain number of \textit{k}-meshes, one will not get any new node points with the further change in size of \textit{k}-mesh. At this stage, it is expected that all the nodes present in the given material have been obtained. Through this approach, in most of the case, all the nodes will be obtained by running the calculations over a smaller number of $\textbf{\textit{k}}$-points in comparison to the previous mentioned method.

\subsection{Technical details}
Entire code is written in Python 3. Thus, the code may not be compatible with the lower versions of python. Different modules of python such as \textit{random}, \textit{math}, \textit{Fraction}, etc., are explicitly used in this code. In addition to these, the \textit{time} module is used to calculate the total time taken in the entire nodes-finding process. The present version of this code is interfaced only with the WIEN2k package. Furthermore, this code can be easily interfaced with the other well-known \textit{first-principle} packages such as elk\cite{elk}. It is important to mention here that the present version of the \textit{PY-Nodes} code is in its most general form. That is, it not designed by considering any specific situation or for any special class of materials. This is to ensure the wide application of the code in the study of any kind of materials hosting the characteristic nodes. 

\begin{table*}
\caption{\label{tab:table1}%
\normalsize{The details of the values assigned to the input parameters for the\textit{ PY-Nodes} code. The size of the \textit{k}-mesh used for finding the nodes is also mentioned. The \textit{k}-meshes are taken in the full Brillouin zone (FBZ).
}}
\begin{ruledtabular}
\begin{tabular}{ccccccccc}
\textrm{$\textbf{\textit{case}}$}&
\textrm{$\textbf{\textit{struct\_num}}$}&
\textrm{$\textbf{coefficients}$}&
\textrm{$\textbf{\textit{SOC}}$}&
\textrm{\textbf{\textit{shift}}}&
\textrm{\textbf{\textit{coordinates\_prec}}}&
\textrm{\textbf{\textit{band\_indices}}}&
\textrm{\textbf{\textit{write\_lim}}}&
\textrm{\textbf{\textit{k}-mesh (FBZ)}}\\
\colrule
         &        &  \textit{alpha}=0.6 &        &        &        &             &             & 7$\times$7$\times$7       \\
    TaAs &   5    &  \textit{beta}=0.5  &    y   &  0.3   &  2     & 84;85       & $10^{-4}$   & \&    \\
         &        &  \textit{gamma}=1.0 &        &        &        &             &             & 8$\times$8$\times$8   \\\\
         &        &  \textit{alpha}=0.6 &        &        &        &             &             &                        \\
Na$_3$Bi &   6    &  \textit{beta}=0.5  &    y   &  0.2   &  3     & 84;85       & $10^{-4}$   & 7$\times$7$\times$3    \\
         &        &  \textit{gamma}=1.0 &        &        &        &             &             &                        \\\\
         &        &  \textit{alpha}=0.6 &        &        &        &             &             &                        \\
   YAuPb &   2    &  \textit{beta}=0.5  &    y   &  0.3   &  6     & 42;43       & $10^{-4}$   & 7$\times$7$\times$7    \\
         &        &  \textit{gamma}=1.0 &        &        &        &             &             &                        \\\\
         &        &  \textit{alpha}=0.6 &        &        &        &             &             &                        \\
  CaAgAs &   6    &  \textit{beta}=0.5  &    n   &  0.2   &  4     & 63;64       & $10^{-4}$   & 5$\times$5$\times$7    \\
         &        &  \textit{gamma}=1.0 &        &        &        &             &             &                        \\\\     
\end{tabular}
\end{ruledtabular}
\end{table*}

\section{Test Cases} 
For benchmarking the \textit{PY-Nodes} code, we have tested it on four different materials: (i) TaAs, a well-known Weyl semimetal\cite{TaAs}, (ii) Na$_3$Bi, a popular Dirac semimetal\cite{Na3Bi}, (iii) YAuPb, a claimed non-trivial topological semimetal\cite{YAuPb_claim} and (iv) CaAgAs, a nodal-line semimetal\cite{CaAgAs}. Since all these materials are associated with characteristic node points, they are suitable for the testing purpose of the present code.

\par Tantalum arsenide (TaAs) is a material that crystallizes in a tetragonal body-centered structure\cite{TaAs-struct}. It has been paid much attention from long time due to its topological semimetallic behaviour\cite{TaAs-semimetal}. The material is well studied and is categorized to be Weyl semimetal. In this regard, the literature survey shows that several studies have been already carried out to find and study the Weyl nodes in TaAs\cite{TaAs-lp}. In these works, the material is found to possess 6 Weyl nodes in the irreducible part of the BZ. This corresponds to a total of 24 Weyl nodes in the full BZ.
\par Na$_3$Bi, which is a well-known Dirac semimetal, possesses hexagonal crystal structure\cite{Na3Bi-struct, Na3Bi1, Liu-ZK}. As mentioned before, the prominent signature of Dirac semimetals is the existence of characteristic Dirac nodes. Several studies have been already performed on this material to investigate its semimetallic behaviours. In these studies, it has been reported that the material possesses two Dirac nodes on the either side of the $\Gamma$-point along the $\boldsymbol{k_3}$ direction. The studies also show that the nodes are located in the vicinity of the $\Gamma$-point.
\par YAuPb is a half-Heusler alloy\cite{HH} which crystallizes in a face-centered cubic crystal system\cite{YAuPb-lp}. The material has been already claimed to be topological semimetal\cite{YAuPb_claim}. Furthermore, the semimetallic properties of the compound have already been extensively investigated\cite{Vivek}. In this work, it has been found that the compound shows several signatures of semimetals such as (i) existence of node points with specific chiralities, (ii) berry-curvature associated with these nodes, and (iii) observation of surface states. It has been reported in the work that several node points are obtained in the vicinity of $\Gamma$-point.
\par CaAgAs crystallizes in a hexagonal pyrochlore-type structure\cite{CaAgAs}. It has been reported that in the absence of SOC, the topmost VB and the bottommost CB touch each other to form a nodal-ring structure. However, in the presence of SOC, these two bands get separated leading to an energy gap of $\Delta$= 73 meV\cite{CaAgAs}. Thus, the compound shows the properties of two different topological classes of material. In the absence of SOC, it behaves as the nodal-line semimetal. However, in the presence of SOC, the material acts as topological insulator\cite{CaAgAs}. As the present work demands the nodal-line characteristic, the material is used in the absence of SOC.
\par It is seen in the above discussions that all these materials possess band touching points (near the Fermi-level). These band touching points are in the form of either node points (\textit{e.g.}, TaAs, YAuPb \& Na$_3$Bi) or nodal-line (\textit{e.g.}, CaAgAs). Thus, it is convincing to use these materials for benchmarking the \textit{PY-Nodes} code.

\begin{table}
\caption{\label{tab:table1}%
\normalsize{Time required for searching nodes for different test cases using the \textit{PY-Nodes} code.
}}
\begin{ruledtabular}
\begin{tabular}{ccc}
\textrm{$\textbf{\textit{case}}$}&
\textrm{\textbf{\textit{k}-mesh (FBZ)}}&
\textrm{\textbf{Time required}}\\
\colrule
    TaAs   & 7$\times$7$\times$7     & 9 hours  28 minutes        \\
           & 8$\times$8$\times$8     & 12 hours 47 minutes       \\\\
 Na$_3$Bi  & 7$\times$7$\times$3     & 10 hours  11 minutes       \\\\
    YAuPb  & 7$\times$7$\times$7     & 6 hours  8 minutes         \\\\
    CaAgAs & 5$\times$5$\times$7     & 6 hours  35 minutes        \\\\
\end{tabular}
\end{ruledtabular}
\end{table}

\subsection{Computational Details}
The ground state energy calculations of each material are carried out using WIEN2k package. The details of the space groups, lattice parameters and the Wyckoff positions of different atoms used for each material are mentioned in Table III. It is well-known that the effect of spin-orbit coupling (SOC) is prominent in materials which are composed of heavy elements. In the present case, all the four compounds are composed of heavy elements. However, CaAgAs shows nodal-line behaviour only in the absence of SOC. Hence, SOC is included only for the case of TaAs, YAuPb and Na$_3$Bi. Furthermore, the PBEsol, which is based on generalised gradient approximation (GGA), is used as an exchange correlation functional in these calculations\cite{PBESol}. The \textit{k}-mesh size used in these computations are also mentioned in Table III corresponding to each material. In these calculations, the \textit{k}-meshes are taken in the irreducible part of the BZ (IBZ). Furthermore, the energy convergence limit for the self-consistent method is set to $10^{-4}$ Ry per unit cell for the calculations corresponding to each material.
\par The \textit{k}-mesh size and the values of other input parameters used for the nodes-finding calculation for each material are presented in Table IV. In these computations, the \textit{k}-meshes are taken in the full BZ (FBZ). It must be noted here that for TaAs, two calculations have been carried out for two different sizes of \textit{k}-mesh as mentioned in the table. The final result is the combined output of both the calculations. This strategy has been already discussed in section III A. Here, it is used to efficiently find out all the nodes present in the material. Moving further, the table IV specifies the values of reflection, contraction and expansion coefficients used in each case. In addition to this, the table mentions the details about the values assigned to parameters such as \textit{struct\_num}, \textit{SOC}, \textit{shift} and \textit{write\_lim} in each case. Furthermore, the band indices corresponding to which the node points are obtained are also mentioned for each material. Besides these, the values for the \textit{coordinates\_prec} parameter taken to obtain the final output are stated for each compound. The \textit{halt\_conv\_limit} for \textit{Nelder-Mead's} iterative method is taken to be $10^{-11}$ for all these cases. It is necessary to mention here that the efficiency of any computational code greatly depends on the time taken by it to perform the required task. The present version of \textit{PY-Nodes} code runs in the serial mode. The time required for searching the nodes for each test case is mentioned in table V. Furthermore, it is already discussed that the Weyl nodes possess specific chirality. So, the chirality of the nodes obtained in the case of TaAs is calculated using \textit{WloopPHI}\cite{WloopPhi} code. It is a well-known python code which is interfaced with the WIEN2k package. 
\par It is important to mention here that the mere existence of node points does not guarantee the presence of Weyl or Dirac points in a material. Thus, the properties of the node points obtained from the present code must be well studied before commenting on their topological behaviour.

\begin{table*}
\caption{\label{tab:table1}%
\normalsize{The result obtained corresponding to TaAs is mentioned below. The primitive and conventional coordinates of the node points along with their chirality are presented in the table. The energy of bands with band-indices 84 \& 85 at these nodes along with the corresponding energy difference are also mentioned, in units of Rydberg (Ry).
}}
\begin{ruledtabular}
\begin{tabular}{ccccccccccc}
\textrm{\textbf{S.No.}}&
\textrm{$\textbf{\textit{k}}_{\textbf{1}}\textbf{(\AA$^{\textbf{-1}}$)}$}&
\textrm{$\textbf{\textit{k}}_{\textbf{2}}\textbf{(\AA$^{\textbf{-1}}$)}$}&
\textrm{$\textbf{\textit{k}}_{\textbf{3}}\textbf{(\AA$^{\textbf{-1}}$)}$}&
\textrm{$\textbf{\textit{k}}_{\textbf{x}}\textbf{(\AA$^{\textbf{-1}}$)}$}&
\textrm{$\textbf{\textit{k}}_{\textbf{y}}\textbf{(\AA$^{\textbf{-1}}$)}$}&
\textrm{$\textbf{\textit{k}}_{\textbf{z}}\textbf{(\AA$^{\textbf{-1}}$)}$}&
\textrm{\textbf{E84 (Ry)}}&
\textrm{\textbf{E85 (Ry)}}&
\textrm{\textbf{Gap (Ry)}}&
\textrm{\textbf{Chirality (\textit{C})}}\\
\colrule
    1 & 0.16  &  0.43 & -0.14 &  0.28 &  0.02 &  0.59 & 0.7364192229 & 0.7364192398 & 1.6926E-08 & +1\\
    2 & 0.14  &  0.45 & -0.16 &  0.28 & -0.02 &  0.59 & 0.7364192229 & 0.7364192400 & 1.7063E-08 & +1\\
    3 & 0.45  &  0.14 & -0.43 & -0.28 &  0.02 &  0.59 & 0.7364192229 & 0.7364192401 & 1.7259E-08 & +1\\ 
    4 & 0.43  &  0.16 & -0.45 & -0.28 & -0.02 &  0.59 & 0.7364192229 & 0.7364192399 & 1.6992E-08 & +1\\
    5 & 0.43  &  0.16 & -0.14 &  0.02 &  0.28 &  0.59 & 0.7364192207 & 0.7364192399 & 1.9199E-08 & -1\\
    6 & 0.14  &  0.45 & -0.43 &  0.02 & -0.28 &  0.59 & 0.7364192231 & 0.7364192401 & 1.6966E-08 & -1\\
    7 & 0.45  &  0.14 & -0.16 & -0.02 &  0.28 &  0.59 & 0.7364192232 & 0.7364192399 & 1.6760E-08 & -1\\
    8 & 0.16  &  0.43 & -0.45 & -0.02 & -0.28 &  0.59 & 0.7364192231 & 0.7364192400 & 1.6924E-08 & -1\\
    9 & -0.43  & -0.16 &  0.45 &  0.28 &  0.02 & -0.59 & 0.7364192080 & 0.7364192360 & 2.8040E-08 & +1\\
    10 & -0.45  & -0.14 &  0.43 &  0.28 & -0.02 & -0.59 & 0.7364192229 & 0.7364192398 & 1.6926E-08 & +1\\
    11 & -0.14  & -0.45 &  0.16 & -0.28 &  0.02 & -0.59 & 0.7364187096 & 0.7364219514 & 3.2417E-06 & +1\\
    12 & -0.16  & -0.43 &  0.14 & -0.28 & -0.02 & -0.59 & 0.7364192230 & 0.7364192402 & 1.7270E-08 & +1\\
    13 & -0.16  & -0.43 &  0.45 &  0.02 &  0.28 & -0.59 & 0.7364192238 & 0.7364192459 & 2.2059E-08 & -1\\
    14 & -0.45  & -0.14 &  0.16 &  0.02 & -0.28 & -0.59 & 0.7364187093 & 0.7364219512 & 3.2418E-06 & -1\\
    15 & -0.14  & -0.45 &  0.43 & -0.02 &  0.28 & -0.59 & 0.7364192231 & 0.7364192400 & 1.6924E-08 & -1\\
    16 & -0.43  & -0.16 &  0.14 & -0.02 & -0.28 & -0.59 & 0.7364192231 & 0.7364192398 & 1.6714E-08 & -1\\
    17 & -0.25  &  0.25 &  0.26 &  0.51 &  0.01 &  0.00 & 0.7363620648 & 0.7363636055 & 1.5407E-06 & +1\\
    18 & -0.26  &  0.26 &  0.25 &  0.51 & -0.01 &  0.00 & 0.7363629657 & 0.7363632963 & 3.3066E-07 & +1\\
    19 &  0.26  & -0.26 & -0.25 & -0.51 &  0.01 &  0.00 & 0.7363631936 & 0.7363632088 & 1.4657E-08 & -1\\
    20 &  0.25  & -0.25 & -0.26 & -0.51 & -0.01 &  0.00 & 0.7363631807 & 0.7363632207 & 3.9960E-08 & -1\\
    21 &  0.25  & -0.25 &  0.26 &  0.01 &  0.51 &  0.00 & 0.7363620647 & 0.7363636054 & 1.5407E-06 & +1\\
    22 & -0.26  &  0.26 & -0.25 &  0.01 & -0.51 &  0.00 & 0.7363356527 & 0.7363795692 & 4.3916E-05 & -1\\
    23 &  0.26  & -0.26 &  0.25 & -0.01 &  0.51 &  0.00 & 0.7363631933 & 0.7363632082 & 1.4905E-08 & +1\\
    24 & -0.25  &  0.25 & -0.26 & -0.01 & -0.51 &  0.00 & 0.7363631807 & 0.7363632207 & 3.9973E-08 & -1\\
\end{tabular}
\end{ruledtabular}
\end{table*}

\subsection{Results and Discussion}
In case of TaAs, the results obtained using the \textit{PY-Nodes} code are mentioned in Table VI. It is seen from the table that 24 nodes are obtained corresponding to full BZ. The table shows the coordinates of these nodes in both the primitive (\boldsymbol{$k_1$},\boldsymbol{$k_2$},\boldsymbol{$k_3$}) and the conventional (\boldsymbol{$k_x$},\boldsymbol{$k_y$},\boldsymbol{$k_z$}) basis. Furthermore, the energy of the topmost VB and the bottommost CB along with the energy gap at these node points are also mentioned. In addition to this, to confirm if the nodes obtained are Weyl nodes, their chiralities are calculated using \textit{WloopPHI}\cite{WloopPhi} code. The chiralities of all the node points obtained are also stated in the table. It is seen from the table that 12 pairs of Weyl nodes having equal and opposite chirality are obtained corresponding to the first BZ. Thus, the sum of chirality across the first BZ is obtained as zero, which is the necessary condition in the case of Weyl semimetal. Grassano \textit{et. al.} have extensively investigated the Weyl nodes in the material\cite{TaAs-coor}. The two nodes which were reported in their work are (0.0078,0.5103,0) \& (0.0198,0.2818,0.5905)\cite{TaAs-coor}. These coordinates are nicely matching with the results obtained in the present work.

\begin{table*}
\caption{\label{tab:table1}%
\normalsize{The result obtained corresponding to Na$_3$Bi is mentioned below. The primitive coordinates of the node points along with their energy are presented in the table. The energy of bands with band-index 84 \& 85 at these nodes along with the corresponding energy difference are also mentioned, in units of Rydberg (Ry).
}}
\begin{ruledtabular}
\begin{tabular}{cccccccc}
\textrm{\textbf{S.No.}}&
\textrm{$\textbf{\textit{k}}_{\textbf{1}}\textbf{(\AA$^{\textbf{-1}}$)}$}&
\textrm{$\textbf{\textit{k}}_{\textbf{2}}\textbf{(\AA$^{\textbf{-1}}$)}$}&
\textrm{$\textbf{\textit{k}}_{\textbf{3}}\textbf{(\AA$^{\textbf{-1}}$)}$}&
\textrm{\textbf{E84 (Ry)}}&
\textrm{\textbf{E85 (Ry)}}&
\textrm{\textbf{Gap (Ry)}}\\
\colrule
    1 &  0.000  &  0.000 &  0.126 & 0.1249697434 & 0.1249697624 & 1.9031E-08\\
    2 &  0.000  &  0.000 & -0.126 & 0.1249697557 & 0.1249697613 & 5.5972E-09\\
\end{tabular}
\end{ruledtabular}
\end{table*}

\par On analysing the material Na$_3$Bi using the \textit{PY-Nodes} code, a pair of nodes are obtained. The details of these nodes are presented in Table VII. The table shows the coordinates in the primitive basis. It is seen from the table that the nodes are situated close to Fermi level, where the Fermi level in Na$_3$Bi is fixed at 0.12452 Ry. Also, the pair of nodes are found to be located on the either sides of $\Gamma$-point, in the $\boldsymbol{k_3}$ direction, as reported in the literature\cite{Na3Bi}. In the works of Liu \textit{et. al.} and Xu \textit{et. al.}, the band-crossing in the material is explored and studied in greater depth\cite{Liu-ZK,WS1}. It has been reported in their works that under the effect of SOC, the topmost VB and the bottommost CB touches each other at (0,0,$\pm$$k_D$), where $k_D$ $\sim$ 0.1 \AA. The coordinates are in the units of primitive lattice vectors. The results obtained in the present work very well match with the coordinates of the nodes reported in the above mentioned works.

\begin{figure}[tbh]
  \begin{center}
    \includegraphics[width=0.85\linewidth, height=5.3cm]{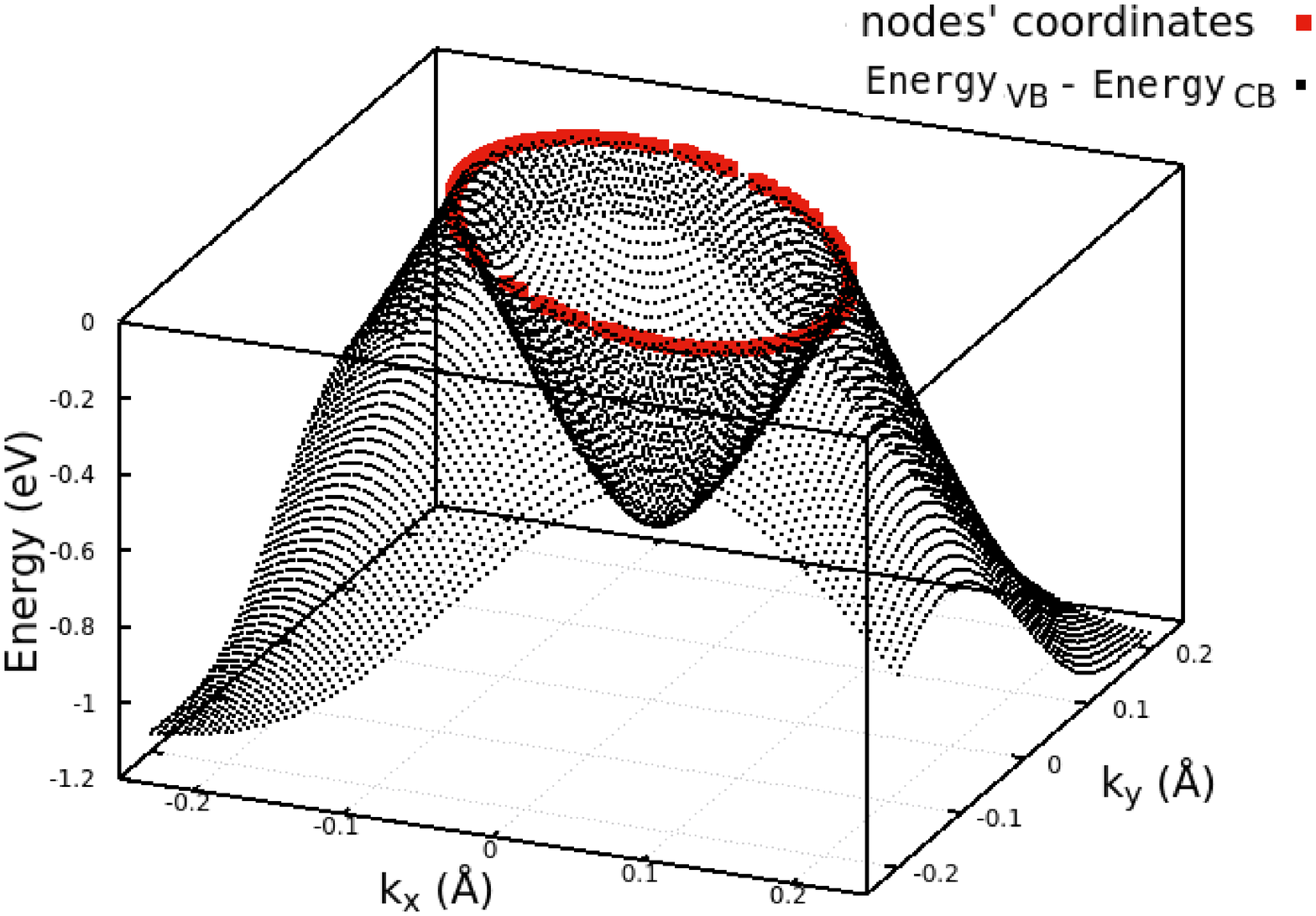} 
    \caption{\small The plot showing the coordinates of nodes in CaAgAs (red) obtained from \textit{PY-Nodes} code getting nicely matched with the region where the term (Energy$_{VB}$ - Energy$_{CB}$) (black) is zero.}
    \label{fig:}
  \end{center}
\end{figure}

\begin{figure}[tbh]
  \begin{center}
    \includegraphics[width=0.85\linewidth, height=5.3cm]{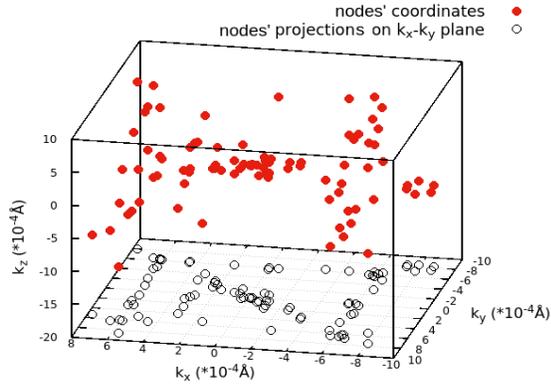} 
    \caption{\small The plot showing the coordinates of nodes in YAuPb obtained from \textit{PY-Nodes} code in the 3D \textit{\textbf{k}}-space (red dots) and their projections on \textit{$k_x$}-\textit{$k_y$} plane (black circles).}
    \label{fig:}
  \end{center}
\end{figure}

\par Moving further to CaAgAs, we have estimated the nodes in the material using the code. The result obtained shows that several nodes are present in the first BZ of the material. All the nodes are obtained in the $k_z$=0 plane. To get a better picture of the position of these nodes, we have plotted the energy gap at these points in the \textit{$k_x$}-\textit{$k_y$} plane. Furthermore, for better visualization, we have also plotted the values of VB's energy - CB's energy in the \textit{$k_x$}-\textit{$k_y$} plane. The corresponding plots are shown in Fig. 2. It is seen from the figure that the values of VB's energy - CB's energy is zero in a circular path in the given coordinate plane. It is further seen that all the nodes obtained from the \textit{PY-Nodes} code sit on this circular path. The nodes are denoted by red dots in the Fig. 2. This circular path which corresponds to the coordinates of the nodes is generally known as nodal-line. Yamakage \textit{et. al.} have performed a detailed analysis of the dispersion curve of CaAgAs. In their work, they found that a nodal-line appears on the $k_z$=0 plane in the absence of SOC\cite{Yamakage}. Furthermore, in another work, the \textit{first-principle} based band-structure calculations have been performed on the material\cite{Takane}. In this study also, the material is reported to possess ring-like nodal-line around the $\Gamma$-point in the absence of SOC. Thus, the results obtained using \textit{PY-Nodes} code are seen to be consistent with the information reported in the literature.

\par Finally, the code has been tested on the material YAuPb. The result obtained shows that a large number of nodes are present in the first BZ of the material. The coordinates of the obtained nodes are plotted which are shown in Fig. 3. The red dots show the coordinates of the nodes in the three-dimensional \textbf{\textit{k}}-space. Also, the black circles show the projections of these nodes on the \textit{$k_x$}-\textit{$k_y$} plane. It is clearly seen from the figure that all the nodes are obtained in the vicinity of the $\Gamma$-point. Pandey \textit{et. al.} have extensively investigated the nodes in YAuPb using the tight-binding approach\cite{Vivek}. In their study, only 10 nodes were obtained. They were found to be concentrated near the $\Gamma$-point. On the other hand, when the material was analysed using the present code, 91 nodes are obtained. All these nodes were found to be situated close to $\Gamma$-point.

\section{Conclusions}

\par In the present work, a Python 3 based code named \textit{PY-Nodes} is designed to search for nodes associated with two or more bands in a given material using the \textit{first-principle} approach. The algorithm of the code is based on the \textit{Nelder-Mead's} function-minimization approach. For benchmarking the code, it has been tested over some well-known materials which possess characteristic nodes. These include - TaAs\cite{TaAs}, Na$_3$Bi\cite{Na3Bi}, CaAgAs\cite{CaAgAs} and YAuPb\cite{YAuPb_claim}. In case of TaAs and Na$_3$Bi, it is seen that the coordinates and the number of nodes obtained are in good match with the details reported in literature\cite{TaAs,Na3Bi}. Specifically, in the case of TaAs, 12 pairs of nodes having equal and opposite non-zero chirality have been obtained. This assures that all the nodes obtained are Weyl nodes. In case of Na$_3$Bi, a pair of nodes are obtained in the vicinity of $\Gamma$-point, along the $\textbf{\textit{k}}_3$ direction. Furthermore, the nodes obtained in case of CaAgAs clearly indicate that the material is a nodal-line semimetal. Finally, when the code is tested on YAuPb, 91 nodes were obtained, all concentrated in a small region close to $\Gamma$-point. These results validate the reliability, efficiency and accuracy of the \textit{PY-Nodes} code for finding out nodes present in a given material. Thus, the code is expected to be very useful for studying the topological systems which possess characteristic node points.

\setlength{\parindent}{3em}
\setlength{\parskip}{0.2em}


\end{document}